\documentclass[aps,prd,eqsecnum,nofootinbib,showpacs,twocolumn]
{revtex4-1}
\usepackage{amsmath,amssymb,bm}
\usepackage[dvips]{graphicx}
\usepackage{color}
\newcommand{\half}{ \frac{1}{2} }
\newcommand{\halfsqrt}{ \frac{1}{\sqrt{2}} }

\newcommand{\hx}{ \Hat{x} }

\newcommand{\ket}[1]{\left|~#1~\right\rangle}

\begin{document}
\title{Closed Superstrings in a Uniform Magnetic Field \\
and \\
Regularization Criterion}
\author{Akira Kokado}
\email{kokado@kobe-kiu.ac.jp}
\affiliation{Kobe International University, Kobe 658-0032, Japan}
\author{Gaku Konisi}
\email{konisigaku@nifty.com}
\affiliation{Department of Physics, Kwansei Gakuin University,
Sanda 669-1337, Japan}
\author{Takesi Saito}
\email{tsaito@k7.dion.ne.jp}
\affiliation{Department of Physics, Kwansei Gakuin University,
Sanda 669-1337, Japan}
\date{\today}
\begin{abstract}
We summarize exact solutions of closed superstrings in a constant magnetic field, from a view point of the regularization criterion. 
Some models will be excluded according to this criterion. The spectrum-generating algebra is also constructed in these interacting models.
\end{abstract}
\maketitle
\section{Introduction}\label{sec:intro}
There is a long history of string models with electromagnetic interactions  \cite{ref:Matsuda_S}-\cite{ref:Kokado_KS}. 
Recently a lot of interest has been drawn in exact solutions of closed strings placed in a uniform magnetic field 
\cite{ref:Russo_A}-\cite{ref:Chizaki_Y}. We can see the Landau-like energy level in these solutions. In these models 
the electromagnetic field has so far been introduced as a Kaluza-Klein type or a gauge field with internal gauge group origin.  \\
\indent In the present paper we first summarize exact solutions of closed superstrings in a uniform magnetic field. 
We give the exact solutions in more generalized forms without taking any light-cone gauge. \\ 
\indent The second aim of this paper is to construct the spectrum-generating algebra (SGA) in these interacting models. Physical states 
satisfying Virasoro conditions or equivalently the BRST charge condition are actually constructed from spectrum-generating operators.\\
\indent As the third aim, we consider a regularization criterion. Some models will be excluded according to this criterion.\\
\indent In Sec.\ref{sec:2} exact solutions of the closed NSR superstring placed in a uniform magnetic field are summarized. 
In Sec.\ref{sec:3} we calculate anomalies associated with the super Virasoro algebra. The detailed calculation of anomalies is 
given by four different methods, which are summarized in Appendices A, B, C and D, in order to emphasize the equivalence between the various regularizations.  
In Sec.\ref{sec:4} the spectrum-generating algebra in our model is constructed.  In Sec.\ref{sec:5} we consider an algebra isomorphic 
to the spectrum-generating algebra. From this we derive the number of space-time dimensions together with normal ordering constants of the 0-th Virasoro operator.  
In Sec.\ref{sec:6} and Sec.\ref{sec:7} exact solutions of the heterotic string in a constant magnetic field are considered. 
We conclude that the heterotic solution is unfortunately excluded according to the regularization criterion. 
The final section is devoted to concluding remarks.  \\
\indent  Appendices are composed of four regularization methods: \\
\indent Appendix A  The operator product expansion method \\
\indent Appendix B  Calculation of anomalies based on contructions \\
\indent Appendix C  Uniqueness of anomalies based on the damping factor method \\
\indent Appendix D  Regularization by means of the generalized zeta function of Riemann  
\section{A closed superstring in a constant magnetic field}\label{sec:2}
The free closed bosonic Lagrangian is given by
\begin{align}
 &  L^0 = -\half \partial_\alpha x_{\mu }\partial^\alpha x^{\mu } 
 = \partial_\alpha x^{+}\partial^\alpha x^{-} - \half \sum_{i=1}^{d-2} \partial_\alpha x_{i}\partial^\alpha x^{i}~,
  \label{eq:Lag1}
\end{align}
where $x^{\pm}=(x^0 \pm x^{d-1})/\sqrt{2}$ are light-cone variables, $x^{d-1}$ being regarded as one of the KK internal coordinates. \\
\indent The electromagnetic field $A_i$ is introduced as the KK type
\begin{align}
 G_{i-}\partial_\alpha x^{-}\partial^\alpha x^{i} + B_{i-}\epsilon _{\alpha \beta }\partial^\alpha x^{-}\partial^\beta x^{i} 
 \nonumber \\
 = A_i \big( \eta _{\alpha \beta } + \epsilon _{\alpha \beta } \big) \partial^\alpha x^{-}\partial^\beta x^{i}~,
 \label{eq:A_KK}
\end{align}
where $G_{i-} = A_i$, and the anti-symmetric background field $B_{i-}=A_i$ is also defined as the same one. Here, two-dimensional matrices are given by
\begin{align}
  \eta _{\alpha \beta }
  = \begin{pmatrix}
              -1 &  0 \\
              0 & 1
    \end{pmatrix}~, \quad
 \epsilon _{\alpha \beta }
  = \begin{pmatrix}
              0 & 1 \\
              -1 & 0
    \end{pmatrix}~.
 \label{eq:def_eta_epsilon}
\end{align}
\indent Introducing right-moving and left-moving variables
\begin{align}
 & s=\tau - \sigma , \quad \partial = \frac{\partial }{\partial s}, \quad \mbox{and} 
 \quad \bar{s} = \tau + \sigma, \quad \bar{\partial}=\frac{\partial}{\partial \bar{s}}~,
 \label{eq:def_s_def}
\end{align} 
we have the total Lagrangian\cite{ref:Russo_A}
\begin{align}
 & L = 2\big[-\bar{\partial }x^{+}\partial x^{-} - \bar{\partial }x^{-}\partial x^{+} 
 + \bar{\partial }x_{i}\partial x^{i} -2A_{i}\bar{\partial }x^{-}\partial x^{i}\big]~.
 \label{eq:TL}
\end{align}
\indent This is extended to the supersymmetric Lagrangian\cite{ref:Russo_A}
\begin{align}
 & \hat{L} = 2\big[-\bar{D}\hat{x}^{+}D\hat{x}^{-} - \bar{D}\hat{x}^{-}D\hat{x}^{+} 
 + \bar{D}\hat{x}_{i}D\hat{x}^{i} 
 \nonumber \\
 & \quad -2A_{i}(\hat{x})\bar{D}\hat{x}^{-}D\hat{x}^{i}\big]~,
 \label{eq:L_2}
\end{align}
where
\begin{align}
 & D=i\partial_{\theta }+ \theta \partial~, \quad \bar{D}=i\partial_{\bar{\theta }}+ \bar{\theta }\bar{\partial}~,
  \label{eq:def_D_barD} \\
 & \hat{x}^{\mu}(s, \bar{s}, \theta, \bar{\theta}) = x^{\mu}(s, \bar{s}) + i \frac{1}{\sqrt{2}}\theta \psi ^{\mu}(s, \bar{s})
  + i \frac{1}{\sqrt{2}}\bar{\theta }\bar{\psi }^{\mu}(s, \bar{s}) 
 \nonumber \\
 & \quad + i\bar{\theta }\theta C^{\mu }(s, \bar{s})~.
  \label{eq:def_x_mu}
\end{align}
We choose the symmetric gauge
\begin{align}
 & A_{i}(\hx ) = - F_{ij} \hx^{j}/2~, \quad F_{ij} = \mbox{const.}~,
 \label{eq:sysgauge}
\end{align}
and concentrate on one of the $2\times 2$ blocks of $F_{ij}$ with $B$ real
\begin{align}
  F_{ij}
  = \begin{pmatrix}
              0 &  B \\
             -B & 0
    \end{pmatrix}~, \quad i, j = 1, 2~,
 \label{eq:defF}
\end{align}
so that
\begin{align}
 & \hat{L} = 2\big[-\bar{D}\hat{x}^{+}D\hat{x}^{-} - \bar{D}\hat{x}^{-}D\hat{x}^{+}   
 \label{eq:L_3} \\
 & \quad + \sum_{i=1}^{2} \big\{\bar{D}\hat{x}_{i}D\hat{x}^{i} 
 + F_{ij}\bar{x}^j\bar{D}\hat{x}^{-}D\hat{x}^{i}\big\}
 + \sum_{k=3}^{d-2} \bar{D}\hat{x}_{k}D\hat{x}^{k} \big]~.
 \nonumber
\end{align}
\indent From this Lagrangian we have the field equations
\begin{align}
 &   \bar{D}D\Hat{x}^{-} = 0~,
 \label{eq:eq_of_motion_x_-} \\
 &  \bar{D}D \hx_{i} + \bar{D}\Hat{x}^{-}F_{ij}D\hx^{j} = 0~, \quad i, j = 1, 2
 \label{eq:eq_of_motion_x_12} \\
  &  \bar{D}D \hx^{+} + \half \bar{D}\big(\Hat{x}\cdot F \cdot D\hx\big) = 0~,
 \label{eq:eq_of_motion_x_+} \\
 & \bar{D}D \Hat{x}^{k} = 0~. \quad k=3, \cdots , d-2
 \label{eq:eq_of_motion_x_k}
\end{align}
Let us solve Eqs.(\ref{eq:eq_of_motion_x_12}) and (\ref{eq:eq_of_motion_x_+}) exactly. Since $\hat{x}^{-}$ satisfies the free field equation 
(\ref{eq:eq_of_motion_x_-}), its solution has a form,
\begin{align}
 & \hat{x}^{-}(s, \bar{s}) = \hat{x}_{R}^{-}(s) + \hat{x}_{L}^{-}(\bar{s})~.
 \label{eq:solv_+-}
\end{align}
By inserting Eq.(\ref{eq:solv_+-}) into Eq.(\ref{eq:eq_of_motion_x_12}), it reduces to
\begin{align}
 &  \bar{D}D \hx  + \bar{D}\Hat{x}_{L}^{-} F\cdot D\hx = 0~.
 \label{eq:eq_of_motion_x_3}
\end{align}
\indent Now, let us define a new variable
\begin{align}
 & \Hat{X}^{i} = \big(\exp{[\Hat{x}_{L}^{-}(\bar{s})F]}\big)^{ij}\hx_{j}~, \quad i, j= 1,2~.
 \label{eq:new_X}
\end{align}
Then we have $D\Hat{X} = \exp{[\Hat{x}_{L}^{-}(\bar{s})F]}\cdot D\hx$, and 
\begin{align}
 & \bar{D}D\Hat{X} = \exp{[\Hat{x}_L^{-}(\bar{s})F]}\cdot [\bar{D}D\hx + \bar{D}\Hat{x}_L^{-} F\cdot D\hx]=0~,
 \label{eq:eq_of_X}
\end{align}
because of Eq.(\ref{eq:eq_of_motion_x_3}). This is nothing but a free equation for $\Hat{X}^{i}$. So we can set as
\begin{align}
 & \hat{X}^{i} (s, \bar{s}) = \Hat{X}_R^{i}(s) + \Hat{X}_L^{i} (\bar{s})~.
 \label{eq:solu_X_KK}
\end{align}
\indent By using Eq.(\ref{eq:solu_X_KK}) the second term in Eq.(\ref{eq:eq_of_motion_x_+}) can be written as
\begin{align}
 & \bar{D}\big( \hx\cdot F \cdot D\hx \big) = \bar{D}\big(\Hat{X}\cdot F \cdot D\Hat{X} \big) 
  \label{eq:formura_X} \\
 & \quad =\bar{D}\Hat{X}\cdot F \cdot D\Hat{X}=\bar{D}\Hat{X}_L\cdot F \cdot D\Hat{X}_R 
  \nonumber \\
 & \quad = \bar{D}D\big(\Hat{X}_L\cdot F \cdot \Hat{X}_R \big) ~.
 \nonumber
\end{align}
Hence, if we define
\begin{align}
 & \Hat{y}^{+} \equiv  \Hat{x}^{+} + \half \big( \Hat{X}_L\cdot F \cdot \Hat{X}_R \big)~,
 \label{eq:def_y}
\end{align}
then Eq.(\ref{eq:eq_of_motion_x_+}) turns out to be
\begin{align}
 & \bar{D}D\Hat{y}^{+} = 0~.
 \label{eq:eq_of_motion_y}
\end{align}
Thus we have found that $\Hat{X}^{1}, \Hat{X}^{2}, \Hat{x}^{-}, \Hat{y}^{+}$ and $\Hat{x}^{k}$ satisfy free field equations. \\
\indent Therefore, we can define an equivalent Lagrangian instead of (\ref{eq:L_3})
\begin{align}
 & \Hat{L}' = 2\big[-\bar{D}\Hat{y}^{+} D\Hat{x}^{-} - \bar{D}\Hat{x}^{-} D\Hat{y}^{+}             
 \label{eq:new_L} \\
 & \quad + \sum_{i=1}^{2}\bar{D}\Hat{X}_{i} D\Hat{X}^{i} 
   + \sum_{k=3}^{d-2}\bar{D}\Hat{x}_{k} D\Hat{x}^{k}\big]~,
 \nonumber
\end{align}
from which free field equations for variables $\Hat{X}^{1}, \Hat{X}^{2}, \Hat{x}^{-}, \Hat{y}^{+}$ and $\Hat{x}^{k}$ are derived. 
When $F_{ij}=0$, the Lagrangian (\ref{eq:new_L}) reduces to the free Lagrangian of (\ref{eq:L_3}). \\
\indent However, in order to solve the field equations from the Lagrangian (\ref{eq:new_L}), it is necessary to find boundary conditions 
for variables at $\sigma =0$ and $2\pi $. We use the complex variable notations, $\Hat{X}^{(\pm)} = (\Hat{X}^1 \pm i\Hat{X}^2)/\sqrt{2}$.  
Then Eq.(\ref{eq:new_X}) turns out to be of the form
\begin{align}
 & \Hat{X}^{(\pm)} = \exp{[\pm i\Hat{x}_L^{-}(\bar{s})B]}\hx^{(\pm)}~.
 \label{eq:refrom_X}
\end{align}
The Lagrangian (\ref{eq:new_L}) can be rewritten as
\begin{align}
 & L' = 2\Big[ \Bar{D}\Hat{X}^{(+)} D\Hat{X}^{(-)} + \Bar{D}\Hat{X}^{(-)} D\Hat{X}^{(+)} 
  - \Bar{D}\Hat{y}^{+} D\Hat{x}^{-}   
 \nonumber \\
 & \qquad  - \Bar{D}\Hat{x}^{-} D\Hat{y}^{+}
  + \sum_{k=3}^{d-2} \Bar{D}\Hat{x}_{k} D\Hat{x}^{k} \big]~.
 \label{eq:L_dash}
\end{align}
Component fields of $\Hat{X}^{(\pm)}$ are defined by
\begin{align}
 & \Hat{X}^{(\pm)} = X^{(\pm)}+ i\frac{1}{\sqrt{2}}\theta \chi ^{(\pm)} 
    + i\frac{1}{\sqrt{2}} \bar{\theta }\bar{\chi }^{(\pm)} + i\bar{\theta }\theta C_X^{(\pm)}~,
 \label{eq:expand_X}
\end{align}
whereas those of $\Hat{x}^{(\pm)}$ and $\Hat{x}_L^{-}$ are
\begin{align}
 & \Hat{x}^{(\pm)} = x^{(\pm)}+ i\frac{1}{\sqrt{2}}\theta \psi ^{(\pm)} 
    + i\frac{1}{\sqrt{2}} \bar{\theta }\bar{\psi }^{(\pm)} + i\bar{\theta }\theta C^{(\pm)}~,
 \label{eq:expand_x} \\
 & \Hat{x}_L^{-} = x_L^{-}+ i\frac{1}{\sqrt{2}}\bar{\theta }\bar{\psi}_{L}^{-} 
    + i\bar{\theta }\theta C_{L}^{-}~.
 \label{eq:expand_x_L}
\end{align}
Comparing both sides of Eq.(\ref{eq:refrom_X}), we have
\begin{align}
 & X^{(\pm)} = \exp{(\pm ix_L^-B)}x^{(\pm)}~,
 \label{eq:rel_X_x} \\
 & \chi ^{(\pm)} = \exp{(\pm ix_L^-B)}\psi ^{(\pm)}~,
 \label{eq:rel_chi_psi} \\
 & \bar{\chi }^{(\pm)} = \exp{(\pm ix_L^-B)}\big( \bar{\psi }^{(\pm)} \pm i \sqrt{2}\bar{\psi }_L^{-}x^{(\pm)}\big)~,
 \label{eq:rel_barchi_barpsi} \\
 & C_X^{(\pm)} = \exp{(\pm ix_L^-B)}\big( C^{(\pm)} \pm \frac{1}{\sqrt{2}}\bar{\psi }_L^{-}\psi ^{(\pm)}\big)~.
 \label{eq:rel_BX_B}
\end{align}
Since $x^{(\pm)}$ and $C^{(\pm)}$ are periodic, $\psi ^{(\pm)}$ and $\bar{\psi }^{(\pm)}$ are anti-periodic (NS) 
or periodic (R), and $x_L^{-}(\bar{s}+2\pi) 
= x_L^{-}(\bar{s})+2\pi (\Tilde{\alpha}_{0} ^-/\sqrt{2})$,  we have boundary conditions 
for the component fields as
\begin{align}
 & X^{(\pm)}(\tau , \sigma +2\pi) =\exp{(\pm 2\pi i \omega)} X^{(\pm)}(\tau, \sigma)~,
  \label{eq:boundarycon_X} \\
 & \chi ^{(\pm)}(\tau , \sigma +2\pi) =\mp \exp{(\pm 2\pi i \omega)} \chi ^{(\pm)}(\tau, \sigma)~,
  \label{eq:boundarycon_chi} \\
 & \bar{\chi }^{(\pm)}(\tau , \sigma +2\pi) =\mp \exp{(\pm 2\pi i \omega)} \bar{\chi }^{(\pm)}(\tau, \sigma)~,
  \label{eq:boundarycon_barchi} \\
 & C_X^{(\pm)}(\tau , \sigma +2\pi) = \exp{(\pm 2\pi i \omega)} C_X^{(\pm)}(\tau, \sigma)~,
  \label{eq:boundarycon_B}
\end{align}
where $-$ and $+$ signs in front of the exponential function stand for NS sector and R sector, respectively,
 and $\omega $ is defined by $B\Tilde{\alpha}_{0}^-/\sqrt{2}=N+\omega $, $N\in  Z$, $0<\omega <1$, 
 called the cyclotron frequency. Here, $\Tilde{\alpha}_{0}^-$ is the zero-mode operator of $x_L^-$, 
 related to its momentum $\Tilde{p}^-$ through $\Tilde{p}^-=\sqrt{2}\Tilde{\alpha}_{0}^-$. We regard it as a c-number, concentrating on one of its eigenspace.\\
\indent Integrating $\Hat{L}'$ over $\theta$, $\Bar{\theta}$, we have
\begin{align}
 & L' = 2\Big[ \Bar{\partial }X^{(+)} \partial X^{(-)} + \Bar{\partial }X^{(-)} \partial X^{(+)} 
  +i\half \big(\chi ^{(+)} \Bar{\partial }\chi ^{(-)} 
 \nonumber \\
 & \quad \quad + \chi ^{(-)} \Bar{\partial }\chi ^{(+)}
  + \Bar{\chi }^{(+)}\partial \Bar{\chi }^{(-)} + \Bar{\chi }^{(-)} \partial \Bar{\chi }^{(+)}\big)\Big]  
  \nonumber \\
 & \quad \quad + 2\sum_{\mu \neq 1,2} \Big[\Bar{\partial }x^{\mu} \partial x_{\mu} 
  + i\half \big( \psi ^{\mu}\bar{\partial }\psi _{\mu} + \bar{\psi }^{\mu}\partial\bar{\psi }_{\mu} \big)\Big]~.
 \label{eq:L_dash2}
\end{align}
Here, we have dropped the $C^{\mu}$ field as usual. In spite of the quasi-periodicity of $X^{(\pm)}$, 
$\chi ^{(\pm)}$ and  $\bar{\chi }^{(\pm)}$, their periodic boundary conditions, which is necessary in the  
variational principle, is guaranteed, because the aperiodic phase factors $\exp{(\pm 2\pi i\omega)}$ are always 
canceled out between the $(+)$ and $(-)$ components in the Lagrangian. \\ 
\indent Equations of motion are all of free type, especially, important things are
\begin{align}
 & \Bar{\partial }\partial X^{(\pm)} = 0~,
 \label{eq:eq_of_motion_X2} \\
 & \Bar{\partial }\chi ^{(\pm)} = 0, \quad \partial \Bar{\chi }^{(\pm)} =0~.
 \label{eq:eq_of_motion_chi2}
\end{align}
Their solutions with boundary conditions (\ref{eq:boundarycon_X}), (\ref{eq:boundarycon_chi}) and 
(\ref{eq:boundarycon_barchi}) are given by
\begin{align}
 & X^{(\pm)}(\tau ,\sigma ) = {X_{R}}^{(\pm)}(s) + {X_{L}}^{(\pm)}(\bar{s}) ~,
  \label{eq:expandX2} \\
 & {X_{R}}^{(\pm )}(s) = i \halfsqrt\sum _{n}\frac{1}{n\pm \omega } \exp{[-i(n\pm \omega )s]} {\alpha _n}^{(\pm)}~,
  \nonumber \\
 & {X_{L}}^{(\pm)}(\bar{s}) = i \halfsqrt\sum _{n}\frac{1}{n\mp \omega } \exp{[-i(n\mp \omega )\bar{s}]} {\tilde {\alpha }_n}^{(\pm)}~,
  \nonumber
\end{align}
\begin{align}
 & \chi ^{(\pm)}(s) = \sum _{r\in Z+\half}{b_r}^{(\pm)}\exp{[-i(r\pm \omega )s}~, \mbox{for NS sector}
  \label{eq:expandX3} \\
 & \chi ^{(\pm)}(s) = \sum _{n\in Z}{d_n}^{(\pm)}\exp{[-i(n\pm \omega )s]}~,
 \nonumber \\
 & \quad \mbox{for Ramond sector}~,
  \nonumber
\end{align}
and
\begin{align}
 & \bar{\chi }^{(\pm)}(\bar{s}) = \sum _{r\in Z+\half}{\tilde {b}_r}^{(\pm)}\exp{[-i(r\pm \omega )\bar{s}]}~, 
  \mbox{for NS sector}
  \label{eq:expandX4} \\
 & \bar{\chi }^{(\pm)}(\bar{s}) = \sum _{n\in Z}{\tilde{d}_n}^{(\pm)}\exp{[-i(n\pm \omega )\bar{s}]}~, 
  \nonumber \\
 & \quad  \mbox{for Ramond sector}~.
  \nonumber
\end{align}
\indent  The conjugate momenta to $X^{(\pm)}(\tau , \sigma )$ are
\begin{align}
 &  P^{(\mp)}= \frac{\partial L'}{\partial \big(\partial _\tau X^{(\pm)}\big )} = \partial X^{(\mp)} + \bar{\partial }X^{(\mp)} 
             = \dot {X}^{(\mp)}~.
 \label{eq:defP}
\end{align}
The quantization is accomplished by setting the commutation rules
\begin{align}
  & \big[\, X^{(\pm)}(\tau , \sigma )\,,~P^{(\mp)}(\tau , \sigma ')\, \big] = 2i\, \pi \delta _{\pm\omega }(\sigma -\sigma ')~,
\label{eq:def-XP-CCR}
\end{align}
and other combinations are zero. Here $\delta _{\pm\omega }(\sigma -\sigma ')$ is the delta function 
with the same quasi-periodicity as $X^{(\pm)}(\tau , \sigma )$  with respect to its argument. 
From these we have commutation relations:
\begin{align}
 & \big[\, {\alpha }_m^{(+)}\,,~{\alpha }_n^{(-)}\, \big] = (m+\omega )\delta _{m+n,0}~, \quad 0<\omega <1
\label{eq:def_alpha=ccR} \\
 & \big[\, {\tilde {\alpha }}_m^{(+)}\,,~{\tilde {\alpha }}_n^{(-)}\, \big] = (m-\omega )\delta _{m+n,0}~,
\label{eq:def-beta-CCR}
\end{align}
and other combinations are zero. Since $0<\omega <1$, $\alpha _m^{(-)}$, $\tilde {\alpha }_m^{(+)}$  are annihilation operators 
for $m>0$, and creation operators for $m\leq 0$, while $\alpha _m^{(+)}$, $\tilde {\alpha }_m^{(-)}$  are creation operators for $m<0$, 
and annihilation operators for $m\geq 0$. \\
\indent As for the fermionic parts, in the same way we get, for right-moving modes
\begin{align}
 & \big\{\, b_r^{(+)}\,,~b_s^{(-)}\, \big\} = \delta _{r+s,0}~, \quad \mbox{others }=0~,
\nonumber \\
 & \big\{\, d_m^{(+)}\,,~d_n^{(-)}\, \big\} = \delta _{m+n,0}~, \quad \mbox{others }=0~.
\label{eq:def-d-CCR}
\end{align}
${b_r}^{(\pm)}$ are annihilation operators for $r>0$, and creation operators for $r<0$. 
The same is true for the left-moving modes.  For the Ramond sector, we need a special care on the 0-modes, so the detail will be discussed in Sec.\ref{sec:3} 
and Appendix \ref{sec:appendA}. \\
\indent The full Virasoro operators are almost the same as those of the free Virasoro operators except for the $(\pm)$  modes, 
\begin{align}
 &  L_n = \half \sum_{m=-\infty }^{+\infty }:\big( \alpha _{-m}^{(+)} \alpha _{n+m}^{(-)} + \alpha _{-m}^{(-)} \alpha _{n+m}^{(+)} \big) :
  \label{eq:VirasoroL_n} \\
 & \quad    +\half \sum_{m=-\infty }^{+\infty } \sum_{\mu ,\nu \neq 1,2} : \eta _{\mu \nu } \alpha _{-m}^{\mu } \alpha _{n+m}^{\nu } :
   \nonumber \\
 & \quad  +\half \sum_{r=-\infty }^{+\infty } \Big( r+\frac{n}{2}-\omega \Big) : b_{-r}^{(+)}b_{n+r}^{(-)} : 
 \nonumber \\
 & \quad +\half \sum_{r=-\infty }^{+\infty } \Big( r+\frac{n}{2}+\omega \Big) : b_{-r}^{(-)}b_{n+r}^{(+)} :
 \nonumber   \\
 & \quad  +\half \sum_{r=-\infty }^{+\infty } \sum_{\mu ,\nu \neq 1,2} \Big( r+\frac{n}{2} \Big) :\eta _{\mu \nu } b_{-r}^{\mu } b_{n+r}^{\nu } :
   \nonumber  \\
 & \quad G_r = \sum_{n=-\infty }^{+\infty } \big( \alpha _{-n}^{(+)}b_{r+n}^{(-)} + \alpha_{-n}^{(-)} b_{r+n}^{(+)} \big) 
 \nonumber \\ 
 & \quad + \sum_{n=-\infty }^{+\infty } \sum_{\mu ,\nu \neq 1,2} \eta _{\mu \nu } \alpha _{-n}^{\mu }b_{r+n}^{\nu }~.
   \nonumber
\end{align}
The left-moving Virasoro operators $\tilde {L}_n, \Tilde {G}_r$ are also the same as above, 
but the mode operators should be replaced by the tilded ones with $(+)\leftrightarrow (-)$, i.e., $\alpha _n^{(\pm)}\rightarrow \tilde{\alpha }_n^{(\mp)}$ 
and $b_r^{(\pm)}\rightarrow \tilde{b}_r^{(\mp)}$.    
\section{Calculation of anomaly} \label{sec:3}
The Lagrangian (\ref{eq:L_dash2}) happens to appear as if it is a free type. However, the dynamical variables $X^{(\pm)}(\tau ,\sigma )$, 
$\chi ^{(\pm)}(\tau ,\sigma )$ and $\bar{\chi }^{(\pm)}(\tau ,\sigma )$  are subject to the quasi-periodicity (\ref{eq:boundarycon_X})-(\ref{eq:boundarycon_barchi}), and this 
causes the inclusion of the cyclotron frequency ƒÖ in the commutators (\ref{eq:def_alpha=ccR}) and (\ref{eq:def-beta-CCR}) for mode operators, 
which are different from those of the completely free case. Considering this fact, we should examine the validity of the super Virasoro algebra 
together with anomalies. Here we use the operator product expansion method for calculation of anomalies. Its detail is given in Appendix A.  
However, we give also other regularization methods in Appendices B, C and D, in order to emphasize the equivalence between the various regularizations. \\
\indent The super Virasoro algebra for NS sector is shown to hold in a form
\begin{align}
 & \big[\, L_{m}\,,~L_{n}\, \big] = (m-n)L_{m+n} + \delta _{m+n,0}A_m~,
\label{eq:L+L+CCR} \\
 & \big[\, L_{m}\,,~G_r\, \big] = \big(\frac{m}{2}-r\big)G_{m+r}~,
\label{eq:L-GrCCR} \\
 & \big\{\, G_r\,,~G_s\, \big\} = 2L_{r+s} + \delta _{r+s,0}B_r~,
\label{eq:GrGrCCR} 
\end{align}
where the anomaly terms are given by
\begin{align}
 &  A_m(\mbox{NS}) = \frac{d}{8}m(m^2-1) + m\omega~,
 \label{eq:Am_NS} \\   
 &  B_r(\mbox{NS}) = \frac{d}{2}\big(r^2-\frac{1}{4}\big) + \omega ~,
 \label{eq:Br_NS} \\
 &  qB = N+\omega , \quad N\in Z, \quad 0<\omega <1~.
 \label{eq:defB_q}
\end{align}
Anomalies for the left-moving part are the same as above, $i.e., \Tilde {A}_m=A_m, \Tilde {B}_r=B_r$. \\
\indent However, we find anomaly terms without the cyclotron frequency for Ramond sector
\begin{align}
 &  A_m(\mbox{Ramond}) = \tilde {A}_m(\mbox{Ramond}) = \frac{d}{8}m^3~,
 \label{eq:Am_Ramond} \\   
 &  B_m(\mbox{Ramond}) = \tilde{B}_m(\mbox{Ramond}) = \frac{d}{2}m^2~.
 \label{eq:Br_Ramond} 
\end{align}
%
\section{Spectrum-generating algebra} \label{sec:4}
The SGA for interacting dimensions $\mu=1,2$, or $(+)$, $(-)$, is characterized by the cyclotron frequency $\omega $. 
We summarize it for the right-moving NS sector:
\begin{align}
 & \big[\, A_m^{(+)}\,,~A_n^{(-)}\, \big] = (m+\omega )\delta _{m+n,0}~, 
 \nonumber \\
 & \big\{\, B_r^{(+)}\,,~B_s^{(-)}\, \big\} = \delta _{r+s,0}~, 
 \nonumber \\
 & \big[\, A_m^i\,,~B_r^j\, \big] = 0~,
 \label{eq:SGA1} \\
 & \big[\, A_m^{(\pm)}\,,~A_n^+\, \big] = (m\pm \omega )A_{m+n}^{(\pm)}~, 
 \nonumber \\
 & \big[\, B_r^{(\pm)}\,,~A_n^+\, \big] = \big( \frac{n}{2} + r \pm \omega \big)B_{r+n}^{(\pm)}~, 
 \nonumber \\
 & \big[\, A_m^{(\pm)}\,,~B_r^+\, \big] = (m\pm \omega )B_{m+r}^{(\pm)}~, 
 \nonumber \\
 & \big\{\, B_r^{(\pm)}\,,~B_s^+\, \big\} = A_{r+s}^{(\pm)}~, 
 \nonumber
\end{align}
\begin{align}
 & \big[\, A_m^{+}\,,~A_n^{+}\, \big] = (m-n )A_{m+n}^{+} + m^3\delta _{m+n,0}~,
 \nonumber \\
 & \big[\, A_m^+\,,~B_r^+\, \big] = \big(\frac{m}{2} - r \big)B_{m+r}^+~,
 \label{eq:SGA2} \\
 & \big\{\, B_r^+\,,~B_s^+\, \big\} = 2A_{r+s}^{+} + 4r^2\delta _{r+s,0}~.
 \nonumber
\end{align}
The sub-algebra (\ref{eq:SGA2}) is completely the same as that in Ref.\cite{ref:Brower_F}, so we omit explicit definitions of $A_n^+$ and $B_r^+$.  
They are composed of free operators with light-cone components.
Any operator in Eqs.(\ref{eq:SGA1}) and (\ref{eq:SGA2}) is commutable with the super Virasoro operator $G_r$. \\
\indent The new operators $A_n^{(\pm)}$ and $B_r^{(\pm)}$ in Eq.(\ref{eq:SGA1}) are defined as 
\begin{align}
 & A_n^{(\pm )} = \frac{1}{2\pi} \int_{-\pi }^{\pi } d\tau  A_n^{(\pm )}(\tau )V^{n\pm \omega }(\tau )~,
 \label{eq:defAmpm} \\
 & \quad A_n^{(\pm)}(\tau ) =  P^{(\pm)} - (n\pm \omega )\chi ^{(\pm)}\psi _{-}~, \nonumber \\
 & B_r^{(\pm)} = \frac{1}{2\pi} \int_{-\pi }^{\pi } d\tau  B_r^{(\pm)}(\tau )V^{r\pm \omega }(\tau )~,
 \label{eq:defBrpm} \\
 & \quad B_r^{(\pm)}(\tau ) = \chi ^{(\pm)}\big( 1-\frac{i}{2} \psi_-\partial _{\tau }\psi _- P_{-}^{-2} \big) P_{-}^{1/2}  
 \nonumber \\
& \quad \quad  - \psi _- P^{(\pm)}P_-^{-1/2}~, \nonumber
\end{align}
where
\begin{align}
 & V(\tau ) = :\exp{[iX_-(\tau )}]:~.  
\end{align}
Here, $X_{-}$, $P_{-}$ and $P^{(\pm)}$, as well as $\chi ^{(\pm)}, \psi _{-}$, are all right-moving operators defined by
\begin{align}
 & X_-(\tau ) = \sqrt{2}x_R^{-}(\tau )=x_- + \tau p_- + i\sum_{n \neq 0} n^{-1} \alpha _n^{-} e^{-in\tau }~, 
  \nonumber  \\
 & P_{\pm}(\tau )=\sqrt{2}\partial _\tau x_R^{\pm}(\tau ) = \sum_n e^{-in\tau }\alpha _n^{\pm}~, 
 \label{eq:mode_expand} \\
 & P^{(\pm)}(\tau )=\sqrt{2}\partial _\tau X_R^{(\pm)}(\tau ) = \sum_n e^{-i(n\pm\omega )\tau }\alpha _n^{(\pm)}~.
 \nonumber
\end{align}
The nude indices $\pm$ denote the light-cone components defined as $X_{\pm}=(X^0 \pm X^{d-1})/\sqrt{2}$. The dressed indices $(\pm)$ of $X^{(\pm)}=(X^1 \pm i X^2)/\sqrt{2}$ should be distinguished 
from the light-cone indices $\pm$. The new definitions for $A_m^{(\pm)}$ and $B_r^{(\pm)}$ reduce to the original ones proposed 
by Brower and Friedmann\cite{ref:Brower_F}, if the cyclotron frequency $\omega $ is set to be zero. \\
\indent The proof of our SGA (\ref{eq:SGA1}) is given by the same method as in Ref.\cite{ref:Kokado_KS2}.
%
\section{Isomorphisms} \label{sec:5}
The algebra (\ref{eq:SGA2}) is similar to the super Virasoro algebra for transverse operators
\begin{align}
 & \big[\, L_m^{T}\,,~L_n^{T}\, \big] = (m-n)L_{m+n}^T + A^{T}(m)\delta _{m+n,0}~,
 \nonumber \\
&  \big[\, L_m^{T}\,,~G_r^{T}\, \big] = \big( \frac{m}{2}-r \big)G_{m+r}^{T}~,
\label{eq:LTGrCCR} \\
&  \big\{\, G_r^{T}\,,~G_s^{T}\, \big\} = 2L_{r+s}^{T} + B^{T}(r)\delta _{r+s,0}~,
 \nonumber 
\end{align}  
where
\begin{align}
 &  A^{T}(m) = \frac{d-2}{8}m(m^2-1) + 2ma + m\omega ~,
 \label{eq:ATm} \\
 &  B^{T}(r) = \frac{d - 2}{2}\big(r^2 - \frac{1}{4} \big) + 2a + \omega ~,
 \label{eq:BTm} \\
 &  qB = N+\omega , \quad N\in Z, \quad 0<\omega <1~.
 \label{eq:Beq2}
\end{align}
Here the superscript $T$ means that the operators are constructed from $L_m, G_r$, 
including only the oscillators with spacial components $\mu =1,2,\cdots, d-2$. The constant $a$ 
is included in $L_{m}^{T}$ as $-a\delta _{m,0}$. \\
\indent  The isomorphisms 
\begin{align}
  A_{m}^{+} \sim L_{m}^{T}, \quad B_{r}^{+} \sim G_r^{T}~,
 \label{eq:AmLTBrGT}
\end{align}
are completed, if there hold equations
\begin{align}
 &  A^{T}(m) = \frac{d-2}{8}(m^3-m) + 2ma + m\omega = m^3~,
 \label{eq:ATm2} \\
 &  B^{T}(r) = \frac{d - 2}{2}\big (r^2 - \frac{1}{4} \big ) + 2a + \omega =4r^2~.
 \nonumber
\end{align}
These two equations are consistent to give the solution,
\begin{align}
 &  d = 10~,
 \label{eq:d-} \\
 &  a = \half (1-\omega )~.
 \nonumber
\end{align}
As for the Ramond sector, we have $d^{R} = 10$ and $a^{R}=0$. \\
\indent  The isomorphisms (\ref{eq:AmLTBrGT}) are also extended to other components interacting 
with the magnetic field. The algebra (\ref{eq:SGA1}) is similar to
\begin{align}
 & \big[\, \alpha _m^{(+)}\,,~\alpha _n^{(-)}\, \big] = (m+\omega )\delta _{m+n,0}~, 
 \nonumber \\
 & \big\{\, b_r^{(+)}\,,~b_s^{(-)}\, \big\} = \delta _{r+s,0}~,  
 \nonumber \\
 & \big[\, \alpha _m^i\,,~b_r^j\, \big] = 0~,
 \nonumber \\
 &  \big[\, \alpha _m^{(\pm)}\,,~L_n^T\, \big] = (m\pm \omega )\alpha _{m+n}^{(\pm)}~, 
 \label{eq:SGA1a} \\
 & \big[\, b_r^{(\pm)}\,,~L_n^T\, \big] = \big( \frac{n}{2} 
+ r \pm \omega \big)b_{r+n}^{(\pm)}~, 
 \nonumber \\
 & \big[\,\alpha _m^{(\pm)}\,,~G_r^T\, \big] = (m\pm \omega )b_{m+r}^{(\pm)}~, 
 \nonumber \\
 &  \big\{\, b_r^{(\pm)}\,,~G_s^T\, \big\} = \alpha _{r+s}^{(\pm)}~. 
 \nonumber 
\end{align}
The isomorphisms are now completed by
\begin{align}
   A_{m}^{(\pm)} \sim \alpha _m^{(\pm)}, \quad B_r^{(\pm)} \sim b_r^{(\pm)}~.
 \label{eq:AmbetaBrbr}
\end{align}
The same conclusion is obtained for the left-moving part. \\
\indent Any physical state should satisfy the BRST condition  $Q_{\mbox{BRST}}\ket{\mbox{phys.}}=0$, 
or equivalently the super Virasoro conditions, 
\begin{align}
 & G_{r>0}\ket{\mbox{phys.}}=0~, \quad \big( L_{n\geq 0}-\delta _{n,0}a \big)\ket{\mbox{phys.}}=0~,
  \label{eq:physcond_Gr_Ln} \\
 & \Tilde {G}_{r>0}\ket{\mbox{phys.}}=0~, \quad \big(\tilde {L}_{n\geq 0}-\delta _{n,0}a\big)\ket{\mbox{phys.}}=0~,
  \label{eq:physcond_tilGr_tilLn}
\end{align}
for the NS sector with $a=(1-\omega )/2$, and 
\begin{align}
 & F_{n\geq 0}\ket{\mbox{phys.}}=0~, \quad L_{n\geq 0}\ket{\mbox{phys.}}=0~,
  \label{eq:physcond_Fr_Ln} \\
 & \Tilde {F}_{n\geq 0}\ket{\mbox{phys.}}=0~, \quad \tilde {L}_{n\geq 0}\ket{\mbox{phys.}}=0~,
  \label{eq:physcond_tilFr_tilLn}
\end{align}
for the Ramond sector. It is well known that such physical states can be constructed by using spectrum-generating operators.
%
%
%
\section{Anomalies in related solvable models} \label{sec:6}
\subsection{A closed bosonic string}
%
If the fermionic field $\psi ^{\mu}(\tau ,\sigma )$ is neglected in our model, we have a closed bosonic string in the constant magnetic field.  
In this case the Virasoro constraint constant is given by $a=1-(\omega -\omega ^2)/2$, and the space-time dimension is $d=26$. The $(\omega -\omega ^2)/2$ 
anomaly comes from the operator product expansion
\begin{align}
 & T^B(z)T^B(z') = \frac{1}{(z-z')^4}+\frac{\omega -\omega^2}{zz'(z-z')^2} 
 \nonumber \\
 & \quad \quad + \frac{2T^B(z')}{(z-z')^2} + \frac{\partial 'T^B(z')}{z-z'}~,
 \label{eq:TBT'B0}
\end{align}
for the bosonic energy-momentum tensor, as is seen by Eq.(\ref{eq:TBT'B}) in Appendix \ref{sec:appendA}.
%
\subsection{The heterotic string}
%
As an interesting possibility of exactly solvable models, we consider the heterotic string [9][10] in the constant magnetic field. This heterotic model is obtained from our model by replacing the left-moving fermion with the 32 Lorentz singlet Majorana-Weyl fermion $\lambda ^A(\tau , \sigma ), A=1, \cdots, 32$.  
The Lagrangian is given by
\begin{align}
 & \hat{L} = 2\Big[-\bar{\partial }\hat{x}^+ D\hat{x}^- - \bar{\partial }\hat{x}^- D\hat{x}^+ 
 \nonumber \\
 & \quad + \sum_{i=1}^{2} \big\{ \bar{\partial }\hat{x}_i D\hat{x}^i + F_{ij}\hat{x}^j \bar{\partial }\hat{x}^- D\hat{x}^i\big\} 
 \label{eq:intro_HL} \\
& \quad + \sum_{k=3}^{d-2}\bar{\partial }\hat{x}_k D\hat{x}^k \Big] + 2i\theta \sum_{A=1}^{32} \lambda^A \partial \lambda^A~.
  \nonumber
\end{align}
The heterotic model is exactly solvable. Unfortunately, however, this heterotic model contains inconsistency coming from anomalies, 
and thereby it fails to be valid. \\
\indent The reason is as follows: The heterotic string in a constant magnetic field is characterized by the Virasoro constraint 
equations, two of which are given by 
\begin{align}
 & \big( L_{0}- a \big)\ket{\mbox{phys.}}=\big(\tilde {L}_{0}-\tilde{a}\big)\ket{\mbox{phys.}}=0~,
  \label{eq:physcond_L0} 
\end{align}
where
\begin{align}
 & a=(1-\omega)/2, \quad \Tilde{a}=1-(\omega -\omega ^2)/2~,
  \label{eq:a_tild_a} \\
 & L_{0}=\frac{\bar{p}^2}{2}+N, \quad \bar{p}^2\equiv \sum_{\mu ,\nu \neq 1,2}\eta _{\mu \nu }p^\mu p^\nu ~,
  \label{eq:cond_L0} \\
 & \Tilde{L}_{0}=\frac{\bar{p}^2}{2}+\tilde{N},
\end{align}
with
\begin{align}
 &  N = \sum_{n=1 }^{\infty }\big( \alpha _{-n}^{(+)} \alpha _{n}^{(-)} + \alpha _{-n}^{(-)} \alpha _{n}^{(+)} \big) 
       + \alpha _{0}^{(-)} \alpha _{0}^{(+)}
 \label{eq:def_N} \\
  & \quad + \sum_{n=1}^{\infty } \sum_{\mu ,\nu \neq 1,2} \eta _{\mu \nu } \alpha _{-n}^{\mu } \alpha _{n}^{\nu } +\sum_{r=1/2 }^{\infty } \big( r-\omega \big) b_{-r}^{(+)}b_{r}^{(-)} 
 \nonumber \\
 & \quad   +\sum_{r=1/2 }^{\infty } \big( r+\omega \big) b_{-r}^{(-)}b_{r}^{(+)} 
+ \sum_{r=1/2}^{\infty } \sum_{\mu ,\nu \neq 1,2} \eta _{\mu \nu } r b_{-r}^{\mu } b_{r}^{\nu } 
  \nonumber  \\
 & \Tilde{N}= \sum_{n=1 }^{\infty }\big( \Tilde{\alpha }_{-n}^{(+)} \Tilde{\alpha }_{n}^{(-)} 
        + \Tilde{\alpha }_{-n}^{(-)} \Tilde{\alpha }_{n}^{(+)} \big) 
        + \Tilde{\alpha }_{0}^{(+)} \Tilde{\alpha }_{0}^{(-)}
 \label{eq:def_tild_N} \\
 & \quad   + \sum_{n=1}^{\infty } \sum_{\mu ,\nu \neq 1,2} \eta _{\mu \nu } \Tilde{\alpha }_{-n}^{\mu } \Tilde{\alpha }_{n}^{\nu }
         + \sum_{A=1}^{32} \sum_{r=1/2}^{\infty } r\lambda _{-r}^A\lambda _{r}^A~.
 \nonumber  
\end{align}
Here we concern the NS superstring for the right-moving sector. The $(\omega -\omega ^2)/2$ anomaly in Eqs.(\ref{eq:a_tild_a}) 
comes from the anomaly term in Eq.(\ref{eq:TBT'B0}). \\
\indent Let us now consider the difference
\begin{align}
 & \big(L_{0}-\tilde {L}_{0}-a+\tilde{a}\big)\ket{\mbox{phys.}}
 \nonumber \\
 & =\big(N-\tilde {N}+\half +\half\omega ^2\big)\ket{\mbox{phys.}}=0~.
  \label{eq:physcond_L0_tild_L0}
\end{align}
The last equation fails to be valid, because the $\omega ^2$ term cannot be canceled by any eigenvalue 
of the number operator difference, $N-\tilde{N}$, which gives at most the first order of $\omega $. 
Therefore, there is no possibility of the solvable heterotic model with Lagrangian (\ref{eq:intro_HL}). \\ 
%
%
%
\section{Uniqueness of regularization}\label{sec:7}
%
In calculation of the normal ordering constant $a$ of the Virasoro operator $L_0$, some authors \cite{ref:Russo_AT2} have proposed 
a new kind of regularization based on the formula
\begin{align}
 & \lim_{e\to 0} \sum_{n=1}^\infty n^{-e}(n +  \omega ) = - \frac{1}{12} -\frac{\omega }{2}~.
 \label{eq:def_c_0}
\end{align}
This differs by $-\omega ^2/2$  from the usual regularization based on the generalized zeta  function of Riemann defined as
\begin{align}
 & \zeta (s, a) = \sum_{n=0}^{\infty } (n+a)^{-s}~, \quad 0 < a \leq 1~,
 \label{eq:zeta_func}
\end{align}
from which we have (see Appendix \ref{sec:appendD})
\begin{align}
 & \zeta (-1, \omega )-\omega  = \lim_{s \to -1} \sum_{n=1}^{\infty } (n+\omega )^{-s} = - \frac{1}{12} -\frac{\omega^2}{2}-\frac{\omega }{2}~.
 \label{eq:zeta_func-1}
\end{align}
Their proposal is based on their observation that the regularization of divergent sum is ambiguous. They have proposed that the correct way to regularize it, 
which leads to a modular invariant partition function, is to use the first prescription (\ref{eq:def_c_0}). \\
\indent If the regularization (\ref{eq:def_c_0}) is used in the heterotic model, the normal ordering constants are given as 
\begin{align}
 & a = \half - \frac{\omega }{2}~, \quad \mbox{for the right-moving NS sector}~,
  \label{eq:regu_a} \\
 & \tilde {a} = 1 - \frac{\omega }{2}~, \quad \mbox{for the left-moving sector}~.
  \label{eq:regu_bar_a}
\end{align}
Since there is no $\omega ^2$ anomaly here, there is no inconsistency in the Virasoro constraint equation,
\begin{align}
 & \big(L_0 - \tilde {L}_{0}-a + \tilde {a}\big)\ket{\mbox{phys.}}
 \nonumber \\
 & = \big(N - \tilde {N} +\half \big)\ket{\mbox{phys.}} =0~.
  \label{eq:phys_state_eq}
\end{align}
\indent On the other hand, the usual regularization (\ref{eq:zeta_func-1}) gives
\begin{align}
 & a = \half - \frac{\omega }{2}~, \quad \mbox{for the right-moving NS sector}~,
  \label{eq:regu_a} \\
 & \tilde {a} = 1 - \frac{\omega }{2} + \frac{\omega^2}{2}~, \quad \mbox{for the left-moving sector}~,
  \label{eq:regu_bar_a}
\end{align}
which lead to the same inconsistency as Eq.(\ref{eq:physcond_L0_tild_L0}). \\
\indent However, we would like to stress that the first regularization prescription (\ref{eq:def_c_0}) is inconsistent with usual regularization prescriptions, 
such as the operator product expansion in Appendix A, the contraction method in Appendix B and the damping factor method in Appendix C. On the other hand, 
the second regularization prescription (\ref{eq:zeta_func}) based on the generalized zeta function of Riemann is consistent with those of 
three regularizations A, B and C. 
The uniqueness of regularization can be specially seen in the generalized damping factor method of Appendix C. 
%
\section{Concluding remarks}\label{sec:11}
%
The heterotic string in a constant magnetic field can be solved exactly for the KK type without taking any light-cone gauge. However, 
we pointed out that they include inconsistency coming from anomalies, which was explicitly explained in Eq.(\ref{eq:physcond_L0_tild_L0}). 
The bosonic string in the left-moving sector carries the anomaly, $(\omega^2 -\omega)/2$, whereas the superstring (NS or R) 
in the right-moving sector carries the anomay, ($-\omega /2$ or 0). The $\omega ^2$ factor in Eq.(\ref{eq:physcond_L0_tild_L0}) cannot be canceled by any eigenvalue 
of the number operator difference, $N-\tilde{N}$, which gives at most the first order of $\omega $.\\
\indent From this observation, we conclude that the exact solution in NSR superstring with the constant magnetic field exists 
in each of combined sectors, (NS-NS), (NS-R) and (R-R), where there is no $\omega ^2$ anomaly. Of course, (bosonic-bosonic) combination 
is allowed to have the exact solution in a constant magnetic field. \\
\indent We have also given the spectrum-generating algebra for our interacting system, which is necessary to construct actually physical states 
satisfying the Virasoro conditions.  
%
\begin{acknowledgments}
\indent We thank T. Okamura for useful discussions.@Thanks are also due to J. G. Russo and E. Kiritsis, who kindly informed us about their early works. 
Especially we owe a lot to J. G. Russo and A. A. Tseytlin for continuous long time (almost one year) discussions and valuable comments.

\end{acknowledgments}
\appendix
\section{Calculation of anomalies based on the operator product expansion method}\label{sec:appendA}
\subsection{The Neveu-Schwarz sector}
It is enough to consider only the right-moving part. Let us define current operators for interacting parts by
\begin{align}
 & J^{(\pm)}(z) = i\partial_{z}X_{R}^{(\pm)}(z) = \halfsqrt z^{\mp\omega } \sum_n z^{-n-1} \alpha _n^{(\pm)} 
 \nonumber \\
 & = \halfsqrt z^{\mp\omega }J_{0}^{(\pm)}(z)~,
 \label{eq:defJ} \\
 & z=\exp{(is)}~,
 \nonumber
\end{align}
with
\begin{align}
 & J_{0}^{(\pm)}(z) = \sum_n z^{-n-1} \alpha _n^{(\pm)}~.
\label{eq:defJ0}
\end{align}  
In the following we use the notation $\partial $ for the derivative $\partial_{z}=\partial /\partial z $ by omitting the index $z$. \\
\indent The operator product expansions for them are given by
\begin{align}
 &  J_{0}^{(+)}(z)J_{0}^{(-)}(z') = \frac{1}{(z-z')^2} + \frac{\omega }{z'(z-z')}~,
 \label{eq:J+J-} \\
 &  J_{0}^{(-)}(z)J_{0}^{(+)}(z') = \frac{1}{(z-z')^2} - \frac{\omega }{z(z-z')}~,
 \label{eq:J-J+} 
\end{align} 
Here, we have used the following contractions:
\begin{align}
 & \langle~\alpha _m^{(+)}\alpha _n^{(-)}~\rangle = \delta _{m+n,0} \theta _{m\geq 0}(m+\omega )~,
  \nonumber \\
 & \langle~\alpha _m^{(-)}\alpha _n^{(+)}~\rangle = \delta _{m+n,0} \theta _{m> 0}(m-\omega )~,
 \label{eq:VEV_alpha_alpha0} \\
 & \quad 0 \leq  \omega < 1~, \quad  
   \theta _\Gamma = \left\{
    \begin{array}{rl}
     1,& \quad \mbox{if $\Gamma$ is true} \\
     0,& \quad \mbox{if $\Gamma$ is false}
    \end{array}\right.
 \nonumber 
\end{align}
For the fermionic fields we confine ourselves to the NS sector,
\begin{align}
 &  \chi ^{(\pm)}(z) = z^{\mp\omega } \sum _r z^{-r-1/2} {b_r}^{(\pm)} = z^{\mp \omega }\chi _{0}^{(\pm)}(z)~,
 \label{eq:def_chi} \\
 &  \chi _{0}^{(\pm)}(z)\chi _{0}^{(\mp)}(z') = \frac{1}{z-z'}~,
 \label{eq:_chi_chi2}
\end{align}
with contractions $\left\langle b_r^{(+)}b_s^{(-)} \right\rangle =\left\langle b_r^{(-)}b_s^{(+)} \right\rangle = \delta _{t+s,0}\theta _{r>0}$ . 
The exponent |1/2 on $z$ in Eq.(\ref{eq:def_chi}) is only for convenience. \\
\indent Define the super current operator for interacting parts by
\begin{align}
 &  G(z) = \sqrt{2} \sum^2_{\mu =1} \chi _{\mu }(z) J^{\mu }(z) 
 \label{eq:defG} \\
 & \qquad = \chi _{0}^{(+)}(z) J_{0}^{(-)}(z) + \chi _{0}^{(-)}(z)J_{0}^{(+)}(z)~.
 \nonumber
\end{align}
Then we calculate the operator product $G(z)G(z')$ to yield the conformal operator $T(z)$, i.e.,
\begin{align}
 &  G(z)G(z') = \frac{2}{(z-z')^3} + \frac{\omega }{zz'(z-z')} + \frac{2T(z')}{z-z'}~,
 \label{eq:GG'}
\end{align}
where
\begin{align}
 &  T(z) = T^B(z) + T^F(z) 
 \label{eq:defT} \\
 & \quad = \half\sum_{\mu =1}^{2}\big[ :J_{0\mu }J_{0}^{\ \mu }: + :\partial \chi _\mu \chi^\mu :\big] ~.
 \nonumber
\end{align}
Here we have used the formula for the fermionic part
\begin{align}
 & :\partial \chi _\mu \chi^\mu : = :\partial \chi^{(+)} \chi^{(-)} + \partial \chi^{(-)} \chi^{(+)} :
 \label{eq:formula_fermion} \\
 & \quad \quad = :\partial\big( z^{-\omega } \chi_{0}^{(+)}\big) z^{+\omega }\chi_{0}^{(-)} + \partial \big( z^{+\omega }\chi_0^{(-)}\big) z^{-\omega }\chi_{0}^{(+)} : 
 \nonumber \\
 & \quad \quad = :- \frac{2\omega }{z}\chi_{0}^{(+)} \chi_{0}^{(-)} + \partial \chi_{0}^{(+)} \chi_{0}^{(-)} 
+ \partial \chi_{0}^{(-)} \chi_{0}^{(+)} : ~. \nonumber
\end{align}
In the same way we get
\begin{align}
 &  T^B(z)T^B(z') = \frac{1}{(z-z')^4} + \frac{\omega - \omega ^2}{zz'(z-z')^2} 
 + \frac{2T^B(z')}{(z-z')^2}
 \nonumber \\
 & \qquad + \frac{\partial 'T^B(z')}{z-z'}~,
 \label{eq:TBT'B}
\end{align}
for the bosonic part  $T^{B} = (1/2)\sum_{\mu =1}^{2}:J_{0\mu }J_{0}^{\mu }:$, and
\begin{align}
 & T^F(z)T^F(z') = \frac{1/2}{(z-z')^4} + \frac{\omega ^2}{zz'(z-z')^2} + \frac{2T^F(z')}{(z-z')^2}
 \nonumber \\
& \qquad + \frac{\partial 'T^F(z')}{z-z'}~,
 \label{eq:TFTF'}
\end{align}
for the fermionic part $T^{F}=(1/2):\partial \chi \cdot\chi :$.  Totally, it follows that
\begin{align}
 &  T(z)T(z') = \frac{3/2}{(z-z')^4} + \frac{\omega }{zz'(z-z')^2} + \frac{2T(z')}{(z-z')^2}
 \nonumber \\
 & \qquad+ \frac{\partial 'T(z')}{z-z'}~.
 \label{eq:TT'2}
\end{align}
It is remarkable that the $\omega ^2$ anomaly in each of the bosonic term in Eq.(\ref{eq:TBT'B}) and the fermionic term in Eq.(\ref{eq:TFTF'}) is canceled out with each other in the total equation in (\ref{eq:TT'2}). The algebra is closed by
\begin{align}
   T(z)G(z') = \frac{3/2}{(z-z')^2}G(z') + \frac{1}{z-z'}\partial 'G(z') ~.
 \label{eq:T-G'}
\end{align}
\indent We have so far considered only the $(1, 2)$ plane, where the constant magnetic field is placed. 
The other ($d-2$) space-time components of fields are all free, and their Virasoro algebras are well known. Collecting all of them, we get
\begin{align}
 &  T(z)T(z') = \frac{3d/4}{(z-z')^4} + \frac{\omega }{zz'(z-z')^2} + \frac{2T(z')}{(z-z')^2}
 \nonumber \\
 & \qquad + \frac{\partial 'T(z')}{z-z'}~,
 \label{eq:T+T+'2} \\
 &  T(z)G(z') = \frac{3/2}{(z-z')^2}G(z') + \frac{1}{z-z'}\partial 'G(z') ~,
 \label{eq:T-G'2} \\
 &  G(z)G(z') = \frac{d}{(z-z')^3} + \frac{\omega }{zz'(z-z')} + \frac{2T(z')}{z-z'}~.
 \label{eq:GG'2} 
\end{align}
These are equivalent to the super Virasoro algebra
\begin{align}
 & \big[\, L_{m}\,,~L_{n}\, \big] = (m-n)L_{m+n} + \delta _{m+n,0}A_m~,
 \nonumber \\
 & \big[\, L_{m}\,,~G_r\, \big] = \big(\frac{m}{2}-r\big)G_{m+r}~,
 \label{eq:L+L+CCR} \\
 & \big\{\, G_r\,,~G_s\, \big\} = 2L_{r+s} + \delta _{r+s,0}B_r~,
\nonumber 
\end{align}
where the anomaly terms are given by
\begin{align}
 &  A_m = \frac{d}{8}m(m^2-1) + m\omega~,
 \label{eq:Am+} \\   
 &  B_r = \frac{d}{2}\big(r^2-\frac{1}{4}\big) + \omega ~,
 \label{eq:Br-} \\
 &  qB = N+\omega , \quad N\in Z, \quad 0<\omega <1~.
 \label{eq:defB_q}
\end{align}
Anomalies for the left-moving part are the same as above, $i.e., \Tilde {A}_m=A_m, \Tilde {B}_r=B_r$. 
%
\subsection{The Ramond sector}
%
As for the Ramond sector, we should be careful for the 0-mode. The mode expansions of fermionic fields are given by
\begin{align}
   \chi _R^{(\pm)}(z) = z^{\mp \omega } \sum_n z^{-n}d_n^{(\pm)} = z^{\mp \omega } \chi _{R0}^{(\pm)}(z)~,
 \label{eq:chi_pm}
\end{align}
where $n$ runs over the integral-numbers. The mode operators obey the commutation relation, 
\begin{align}
  \big\{\, d_m^{(+)}\,,~d_n^{(-)}\, \big\} = \delta _{m+n,0}~.
\label{eq:d-CCR2}
\end{align}
Usually the 0-mode $d_0^{\mu }$ is regarded as the Dirac  $\gamma $-matrix. However, in the presence of the magnetic field, 
it is not the case. The reason is as follows: Note that the super Virasoro operator $F_0$ contains factors, 
$\alpha _0^{(+)}d_0^{(-)}+\alpha _0^{(-)}d_0^{(+)}$. Since $\alpha _0^{(-)}$ is the creation operator, 
the second term contradicts with the Virasoro condition $F_0\ket{\mbox{ground state}}=0$, if $d_0^{(\pm)}$ is regarded as the Dirac  $\gamma $ matrix. 
In the sector of the presence of magnetic fields, therefore, $d_0^{(+)}$ should be regarded as the annihilation operator, 
whereas other components $d_0^{\mu }$ without magnetic fields behave as $\gamma $ matrices. \\
\indent From this reason $d_m^{(+)}$ is regarded as annihilation operator for $m\geq 0$, 
and creation operator for $m<0$, while $d_m^{(-)}$ is annihilation operator for $m>0$, and creation operator for $m\leq 0$. 
The contractions are, therefore, defined as
\begin{align}
 & \left\langle~d_m^{(+)}d_n^{(-)} ~\right\rangle =
  \left\{\begin{array}{rl}
        \delta _{m+n,0},& \quad (m\geq 0) \\
        0,& \quad (m<0)
         \end{array}\right.  
 \nonumber \\
 & \left\langle~d_m^{(-)}d_n^{(+)} ~\right\rangle =
  \left\{\begin{array}{rl}
        \delta _{m+n,0},& \quad (m>0) \\
        0.& \quad (m\leq 0)
         \end{array}\right.
 \label{eq:d_d} 
\end{align}
\\ 
The operator product expansions for fermionic fields are, then, given by
\begin{align}
 & {\chi }_{R0}^{(+)}(z){\chi }_{R0}^{(-)}(z') = \frac{z}{z-z'}~,
 \label{eq:chi+chi-2} \\
 & \chi _{R0}^{(-)}(z)\chi _{R0}^{(+)}(z') = \frac{z'}{z-z'}~.
 \label{eq:chi-chi+2}
\end{align}
For the super operator, $F(z)=\chi _{R0}^{(+)}(z)J_0^{(-)}(z)+\chi _{R0}^{(-)}(z)J_0^{(+)}(z)$, 
we have 
\begin{align}
  \mbox{Anomaly terms of } F(z)F(z')=\frac{z+z'}{(z-z')^3}~.
 \nonumber
\end{align}
From the formula
\begin{align}
  \big\{\, F_m\,,~F_n\, \big\} = \oint dz dz'~z^{m}{z'}^{n} F(z)F(z')~,
\label{eq:defFmFn}
\end{align}
it follows that
\begin{align}
 & \big\{\, F_m\,,~F_n\, \big\} = 2L_{m+n} + \delta _{m+n,0}B_m(\mbox{Ramond})~,
\label{eq:FmFn2} \\
 & B_m(\mbox{Ramond}) = \frac{d}{2}m^2~.
\end{align}
For the fermionic part $T^F=(1/2):\partial \chi _R\cdot \chi _R:$ with
\begin{align}
 & 2 T^F = :\partial {\chi _R}^{(+)}{\chi _R}^{(-)} + \partial {\chi _R}^{(-)}{\chi _R}^{(+)} : 
 \nonumber \\
 & = :-\frac{2\omega }{z} {\chi _{R0}}^{(+)}{\chi _{R0}}^{(-)} 
  + \partial {\chi _{R0}}^{(+)}{\chi _{R0}}^{(-)} 
 \nonumber \\
& + \partial {\chi _{R0}}^{(-)}{\chi _{R0}}^{(+)} :~,
 \nonumber
\end{align}
we have
\begin{align}
 & \mbox{Anomaly parts of }T^F(z)T^F(z') = \frac{1}{(z-z')^4}\frac{z^2+z'^2}{4} 
 \nonumber \\
 & \qquad + \frac{\omega ^2 - \omega }{(z-z')^2}~.
 \label{eq:anomalyTFTF'}
\end{align}
\indent For the bosonic part $T^B=:J_0^{(+)}J_0^{(-)}:$, we already had the product $T^B(z)T^B(z')$ before as
\begin{align}
  \mbox{Anomaly parts of} T^B(z)T^B(z') = \frac{zz'}{(z-z')^4} + \frac{\omega -\omega ^2}{(z-z')^2}~.
 \label{eq:anomalyTBTB}
\end{align}  
Here the equation has been multiplied by the factor $zz'$, in order to make it of the same power as the fermionic one. 
Then the total sum of the anomaly is given by
\begin{align}
 & \mbox{Anomaly of } \big[T^B(z) + T^F(z)\big]\big[T^B(z') + T^F(z')\big] 
 \nonumber \\ 
 &   = \frac{1}{(z-z')^4}\big( zz' + \frac{z^2 + z'^2}{4} \big)~,
 \label{eq:anomalyT+T}
\end{align}  
where $\omega, \omega^2$ terms are cancelled out from Eqs.(\ref{eq:anomalyTFTF'}) and (\ref{eq:anomalyTBTB}). 
This gives the anomaly term without the cyclotron frequency
\begin{align}
   A_m(\mbox{Ramond}) = \frac{d}{8}m^3~,
 \label{eq:Am2} \\
   B_m(\mbox{Ramond}) = \frac{d}{2}m^2~.
 \label{eq:Bm2}
\end{align}
The same is true for the left-moving part.
%
%
\section{Calculation of anomalies based on contractions}\label{sec:appendB}
%
The most simple method to obtain anomalies for relevant parts is to calculate contractions of $\big[\, L_m\,,~L_n\, \big]$, or 
$\big\{\, G_r\,,~G_s\, \big\}$.
 For the bosonic case we have
\begin{align}
 & \langle~\big[\, L_m\,,~L_n\, \big]~\rangle = \sum_{k,l} \langle~:\alpha _k^{(+)}\alpha _{m-k}^{(-)}::\alpha _l^{(+)}\alpha _{n-l}^{(-)}: 
 \nonumber \\
 & \qquad - :\alpha _l^{(+)}\alpha _{n-l}^{(-)}::\alpha _k^{(+)}\alpha _{m-k}^{(-)}:~\rangle
 \label{eq:VEV_LL} \\
 & \quad = \sum_{k,l} \langle~\alpha _k^{(+)}\alpha _{n-l}^{(-)}~\rangle \langle~\alpha _{m-k}^{(-)}\alpha _{l}^{(+)}~\rangle 
 \nonumber \\
 & \qquad - \sum_{k,l} \langle~\alpha _l^{(+)}\alpha _{m-k}^{(-)}~\rangle \langle~\alpha _{n-l}^{(-)}\alpha _{k}^{(+)}~\rangle~,
 \nonumber 
\end{align}
where contractions are defined as
\begin{align}
 & \langle~\alpha _m^{(+)}\alpha _n^{(-)}~\rangle = \delta _{m+n,0} \theta _{m\geq 0}(m+\omega )~,
  \nonumber \\
 & \langle~\alpha _m^{(-)}\alpha _n^{(+)}~\rangle = \delta _{m+n,0} \theta _{m> 0}(m-\omega )~,
 \label{eq:VEV_alpha_alpha} \\
 & \quad 0 < \omega < 1~, \quad  
   \theta _\Gamma = \left\{
    \begin{array}{rl}
     1,& \quad \mbox{if $\Gamma$ is true} \\
     0,& \quad \mbox{if $\Gamma$ is false}
    \end{array}\right.
 \nonumber 
\end{align}  
These equations give a finite sum so that we have the unique anomaly $A_m^B$ with $\omega ^2$ term.
\begin{align}
 & \langle~\big[\, L_m\,,~L_n\, \big]~\rangle = \delta _{m+n,0} A_m^B~,
 \label{eq:VEV_LL2} \\
 & \quad A_m^B = \sum_{k=0}^{m-1} ( m-k-\omega )(k+\omega ) 
 \nonumber \\
 & \qquad = \frac{1}{6}m(m^2-1) +m\omega (1-\omega )~.
 \nonumber
\end{align}
\indent In the same way, for the superstring case we have
\begin{align}
 & \langle~G_r~G_s~\rangle = \sum_{m,n} \langle~b_{r-m}^{(+)}\alpha _{m}^{(-)}b_{s-n}^{(-)}\alpha _{n}^{(+)} 
 + b_{r-m}^{(-)}\alpha _{m}^{(+)}b_{s-n}^{(+)}\alpha _{n}^{(-)}~\rangle
 \nonumber \\
 & \quad  = \sum_{m,n} \big(\langle~b_{r-m}^{(+)}\alpha _{s-n}^{(-)}~\rangle \langle~\alpha _{m}^{(-)}\alpha _{n}^{(+)}~\rangle 
   \label{eq:VEV_GG}  \\
 & \qquad + \langle~b_{r-m}^{(-)}b_{s-n}^{(+)}~\rangle \langle~\alpha _{m}^{(+)}\alpha _{n}^{(-)}~\rangle \big)~.
 \nonumber
\end{align}
Here contractions for fermionic operators have been defined as
where contractions are defined as
\begin{align}
 & \langle~b_{r-m}^{(+)}b_{s-n}^{(-)}~\rangle = \langle~b_{r-m}^{(-)}b_{s-n}^{(+)}~\rangle = \delta _{t+s,m+n} \theta _{r-m>0}~,
  \label{eq:VEVbb} 
\end{align}  
so that we get a finite sum for each contraction to yield
\begin{align}
 & \langle~G_r~G_s~\rangle = \delta _{r+s,0}B_r~,
 \label{eq:VEV_GG2} \\
 & \quad B_r = r^2 - \frac{1}{4} + \omega ~.
 \nonumber
\end{align}
The corresponding anomaly of $\big[\, L_m\,,~L_n\, \big]$ will be obtained by using formulae,
\begin{align}
 & \big[\, L_m\,,~L_{r+s}\, \big] = \half \big[\, L_m\,,~\big\{\, G_r\,,~G_s\, \big\}\, \big] ~,
 \label{eq:commutation_LL}
\end{align}
and
\begin{align}
 & \big[\, L_m\,,~G_{r}\, \big] = \big(\half m - r\big) G_{m+r}~. 
 \label{eq:commutation_LG}
\end{align}
The result is
\begin{align}
 & A_m = \frac{1}{4}m\big(m^2 - 1\big)+m\omega ~, 
 \label{eq:A_m_s}
\end{align}
without $\omega ^2$ term.
%
\section{Uniqueness of anomalies based on the damping factor method}\label{sec:appendC}
Commutation relations for bosonic string are given by
\begin{align}
 & \big[\, \alpha _m^{(+)}\,,~\alpha _n^{(-)}\, \big] = (m+\omega )\delta _{m+n,0} g_{m}~, 
 \label{eq:commutation_alpha_alpha_B} \\
 & \big[\, \alpha _m^{(-)}\,,~\alpha _n^{(+)}\, \big] = (m-\omega )\delta _{m+n,0} g_{-m}~,
 \nonumber
\end{align}
where a damping factor $g_m$ is inserted. Its detailed functional form is irrelevant, 
but it should be set as $g_m=1$ when series is a finite sum. \\
\indent The relevant two dimensional Virasoro operator is given by
\begin{align}
 &  L_n = \half \sum_{k}:\big( \alpha _{k}^{(+)} \alpha _{n-k}^{(-)} + \alpha _{k}^{(-)} \alpha _{n-k}^{(+)} \big) : 
   =  \sum_{k} :\alpha _{k}^{(+)} \alpha _{n-k}^{(-)}: ~.
  \label{eq:L_nB}
\end{align}
Then we have
\begin{align}
 & \big[\, L_m\,,~L_{-n}\, \big] = \sum_{k,l}\big[ \alpha _{k}^{(+)} \alpha _{m-k}^{(-)} \,,~
		\alpha _{l}^{(+)} \alpha _{n-l}^{(-)} \, \big]
   \nonumber \\
 & \quad = \sum_{k,l}\Big\{\big[ \alpha _{m-k}^{(-)} \,,~\alpha _{l}^{(+)}\, \big]\alpha _{k}^{(+)} \alpha _{n-l}^{(-)} 
   \nonumber \\
 & \qquad   + \big[ \alpha _{k}^{(+)} \,,~\alpha _{n-l}^{(-)} \, \big]\alpha _{l}^{(+)} \alpha _{m-k}^{(-)} \Big\} 
   \label{eq:commutaion_LL_B} \\
 & \quad = \sum_{k,l} \big\{ \delta _{m-k+l,0}(m-k-\omega )g_{-m+k}\alpha _k^{(+)} \alpha _{n-l}^{(-)} 
 \nonumber \\
 & \qquad + \delta _{k+n-l,0}(k+\omega )g_{k}\alpha _l^{(+)} \alpha _{m-k}^{(-)} \big\} 
 \nonumber \\
 & \quad = \sum_k \big\{ (m-k-\omega )g_{-m+k}\alpha _k^{(+)} \alpha _{m+n-k}^{(-)} 
 \nonumber \\ 
 & \qquad + (k+\omega )g_{k}\alpha _{n+k}^{(+)} \alpha _{m-k}^{(-)} \big\} 
 \nonumber \\
 & \quad = (m-n)L_{m+n} + \delta _{m+n,0}A_m^B~,
 \nonumber 
\end{align}
\begin{align}
 &  A_m^B = \sum _{k\geq 0} (m-k-\omega )(k+\omega )g_{-m+k} g_k 
 \nonumber \\
 & \qquad + \sum _{k\geq m} (k+\omega )(k-m+\omega )g_k g_{k-m}~.
 \nonumber
\end{align}
When $m>0$, it follows that
\begin{align}
 & A_m^B = \sum _{0\leq  k < m} (m-k-\omega )(k+\omega )g_k g_{k-m}
 \nonumber \\
 & \quad = \sum _{k=0}^{m-1} (m-k-\omega )(k+\omega )
 \label{eq:A_m^B_Value} \\
 & \quad = \frac{1}{6}m(m^2-1) + m\omega (1-\omega )~.
 \nonumber
\end{align}
Here the damping factors, $g_k, g_{k-m}$, have been set as unity, since the series is finite. The same is true when $m\leq 0$. 
The result agrees with Eq.(\ref{eq:VEV_LL2}) with $\omega ^2$ term. \\
\indent For the NS string case, commutation relations of fermionic mode operators are given by
\begin{align}
 & \big\{\, b_r^{(+)}\,,~b_s^{(-)}\, \big\} = \delta _{r+s,0} \gamma _{r}~, 
 \quad 
 \big\{\, b_r^{(-)}\,,~b_s^{(+)}\, \big\} = \delta _{r+s,0} \gamma _{-r}~.
 \label{eq:commutation_b_b_B}
\end{align}
where $\gamma _{r}$ is the damping factor. By using Eqs.(\ref{eq:commutation_alpha_alpha_B}) and (\ref{eq:commutation_b_b_B}), let us calculate the anti-commutator
\begin{align}
 & \big\{\, G_r\,,~G_{s}\, \big\} 
 \nonumber \\
 & = \sum_{m,n}\big\{ b_{r-m}^{(+)} \alpha _{m}^{(-)}+b_{r-m}^{(-)} \alpha _{m}^{(+)} \,,~
		b_{s-n}^{(+)} \alpha _{n}^{(-)}+b_{s-n}^{(-)} \alpha _{n}^{(+)} \, \big\}
   \nonumber \\
 &  = \sum_{m,n}\Big[\big\{ b_{r-m}^{(+)} \alpha _{m}^{(-)} \,,~b_{s-n}^{(-)} \alpha _{n}^{(+)}\, \big\}
 \nonumber \\
 & \quad   + \big\{ b_{r-m}^{(-)} \alpha _{m}^{(+)} \,,~b_{s-n}^{(+)} \alpha _{n}^{(-)} \, \big\}\Big] 
   \nonumber \\
 &  = \sum_{m,n}\Big[\big\{ b_{r-m}^{(+)}  \,,~b_{s-n}^{(-)} \, \big\}\alpha _{m}^{(-)}\alpha _{n}^{(+)}
		+\big[ \alpha _{n}^{(+)} \,,~ \alpha _{m}^{(-)}\, \big] b_{s-n}^{(-)}b_{r-m}^{(+)} 
 \nonumber \\
 & \quad  + \big\{ b_{r-m}^{(-)}  \,,~b_{s-n}^{(+)} \, \big\}\alpha _{m}^{(+)}\alpha _{n}^{(-)} 
          +\big[ \alpha _{n}^{(-)} \,,~ \alpha _{m}^{(+)} \, \big]b_{s-n}^{(+)}b_{r-m}^{(-)} \Big] 
  \label{eq:commutaion_G_G_B} \\
 & = \sum_{m,n}\big[ \delta _{r+s-m-n,0}\gamma _{r-m}\alpha _m^{(-)} \alpha _{n}^{(+)} 
  \nonumber \\
 & \quad + \delta _{m+n,0}(n+\omega )g_n b_{s-n}^{(-)}b_{r-m}^{(+)}
    \nonumber \\
 &  \quad + \delta _{r+s-m-n,0}\gamma _{-r+m}\alpha _m^{(+)} \alpha _{n}^{(-)} 
 \nonumber \\
 & \quad + \delta _{m+n,0} (n-\omega )g_{-n} b_{s-n}^{(+)}b_{r-m}^{(-)} \big\} 
  \nonumber \\
 &  = \sum_{m}\big[ \gamma _{r-m}\alpha _m^{(-)} \alpha _{r+s-m}^{(+)}
		+ (-m+\omega )g_{-m} b_{s+m}^{(-)}b_{r-m}^{(+)}
    \nonumber \\
 &  \quad + \gamma _{-r+m}\alpha _m^{(+)} \alpha _{r+s-m}^{(-)} 
        + (-m-\omega )g_{m} b_{s+m}^{(+)}b_{r-m}^{(-)} \big]~. 
  \nonumber
\end{align}
Accordingly we get
\begin{align}
 & \big\{\, G_r\,,~G_{s}\, \big\} = 2L_{r+s} + \delta _{r+s,0}B_r~,
   \label{eq:commution_G_GB2} \\
 & \quad B_r = \sum_{m>0}\gamma _{r-m}g_{-m}(m-\omega ) - \sum_{m>r}\gamma _{r-m}g_{-m}(m-\omega )
   \nonumber \\
 & \quad \quad + \sum_{m\geq 0}\gamma _{-r+m}g_{m}(m+\omega ) - \sum_{m>r}\gamma _{-r+m}g_{m}(m+\omega )~.
   \nonumber
\end{align}
When $r>0$, it follows that
\begin{align}
 & B_r = \sum_{m=1}^{r-1/2} \gamma _{r-m}g_{-m}(m-\omega ) + \sum_{m=0}^{r-1/2} \gamma _{-r+m}g_{m}(m+\omega )~.
   \label{eq:Cal_B_r}
\end{align}
Since this is a finite sum, one can set as $\gamma _r= g_m=1$ to yield 
\begin{align}
 & B_r = 2\sum_{m=1}^{r-1/2} m+\omega=r^2 - \frac{1}{4} +\omega ~.
   \label{eq:Cal_B_r2}
\end{align}
The same is true even when $r<0$. \\
\indent In this calculation we do not need any symmetry of $\gamma _r$, $g_m$.with respect to $\pm r$ and $\pm m$. \\
\indent In conclusion this regularization never depends on any functional form of the damping factor. 
Therefore, there is no ambiguity in this method, giving a unique result.
%
\section{Regularization by means of the generalized zeta function of Riemann}\label{sec:appendD}
The generalized zeta function of Riemann is defined as
\begin{align}
 & \zeta (s, a) = \sum_{n=0}^{\infty } \frac{1}{(n+a)^{s}}~, \quad 0 < a \leq 1~.
 \label{eq:zeta_C}
\end{align}
Especially we have
\begin{align}
 & \zeta (-1, \omega )-\omega  = \sum_{n=1}^{\infty } (n+\omega )=-\frac{1}{12}-\frac{\omega ^2}{2} - \frac{\omega }{2}~, 
 \label{eq:zeta_omega}
\end{align}
and
\begin{align}
 & \sum_{n=1}^{\infty } (n-\omega )=-\frac{1}{12}-\frac{\omega ^2}{2} + \frac{\omega }{2}~, 
 \label{eq:zeta_omega2}
\end{align}
in the region $0 < \omega  < 1$. \\ 
\indent By using these formulae let us calculate normal ordering constants for NS sector. The un-normal ordered Virasoro 0-th operator is given by
\begin{align}
 &  L_0 = \half \sum_{n}\big( \alpha _{-n}^{(+)} \alpha _{n}^{(-)} + \alpha _{-n}^{(-)} \alpha _{n}^{(+)} \big)
 \label{eq:def_L_n_C} \\
 & \quad   +\half \sum_{n} \sum_{\mu ,\nu \neq 1,2} \eta _{\mu \nu } \alpha _{-n}^{\mu } \alpha _{n}^{\nu }
 +\half \sum_{r} \big( r-\omega \big)  b_{-r}^{(+)}b_{r}^{(-)}  
 \nonumber   \\
 &  \quad  
        +\half \sum_{r}\big( r+\omega \big) b_{-r}^{(-)}b_{r}^{(+)} 
        +\half \sum_{r}\sum_{\mu ,\nu \neq 1,2} \eta _{\mu \nu } r b_{-r}^{\mu } b_{r}^{\nu }~,
 \nonumber \
\end{align}
with
\begin{align}
 & \big[\, {\alpha }_n^{(\pm)}\,,~{\alpha }_{-n}^{(\mp)}\, \big] = (n \pm \omega )~,
\label{eq:2}  \\
 & \big\{\, b_r^{(+)}\,,~b_{s}^{(-)}\, \big\} = \delta _{r+s,0}~.
\end{align}
%
\subsection{Bosonic sector}
%
The bosonic sector becomes
\begin{align}
 &  L_0^B = \half \sum_{n\geq 1}\big( \alpha _{-n}^{(+)} \alpha _{n}^{(-)} + \alpha _{-n}^{(-)} \alpha _{n}^{(+)} \big) 
 \label{eq:L_0_B2} \\
 & \quad + \half \sum_{n\geq 1}\big( \alpha _{n}^{(+)} \alpha _{-n}^{(-)} + \alpha _{n}^{(-)} \alpha _{-n}^{(+)} \big) 
  \nonumber  \\
 &  \quad   +\half \sum_{n\geq 1} \sum_{\mu ,\nu \neq 1,2} \eta _{\mu \nu }\big( \alpha _{-n}^{\mu } \alpha _{n}^{\nu } + \alpha _{n}^{\mu } \alpha _{-n}^{\nu } \big) 
 \nonumber \\
 & \quad + \half \big( \alpha _{0}^{(+)} \alpha _{0}^{(-)} + \alpha _{0}^{(-)} \alpha _{0}^{(+)} \big)    
   	    + \half \sum_{\mu ,\nu \neq 1,2} \eta _{\mu \nu } p^{\mu } p^{\nu }
  \nonumber \\
 & \quad = \sum_{n\geq 0}\big( \alpha _{-n}^{(+)} \alpha _{n}^{(-)} + \alpha _{-n}^{(-)} \alpha _{n}^{(+)} \big)
 \nonumber \\
& \quad	+ \half \sum_{n\geq 1} (n+\omega ) + \half \sum_{n\geq 1} (n-\omega ) 
  \nonumber \\
 & \quad 
	    + \sum_{n\geq 1} \sum_{\mu ,\nu \neq 1,2} \eta _{\mu \nu } \alpha _{-n}^{\mu } \alpha _{n}^{\nu }
        + \half\sum_{n\geq 1} \sum_{\mu ,\nu \neq 1,2} \eta _{\mu \nu } \eta ^{\mu \nu } n 
   \nonumber \\
 &  \quad	+ \alpha _{0}^{(-)} \alpha _{0}^{(+)} +\half \omega + \half {p'}^2~.
   \nonumber
\end{align}
From Eqs.(\ref{eq:zeta_omega}) and (\ref{eq:zeta_omega2}) we have
\begin{align}
 & \half \sum_{n\geq 1} (n+\omega ) + \half \sum_{n\geq 1} (n-\omega ) = - \frac{1}{12} - \half \omega ^2~.
  \nonumber
\end{align}
Also
\begin{align}
 & \half\sum_{n\geq 1} \sum_{\mu ,\nu \neq 1,2} \eta _{\mu \nu } \eta ^{\mu \nu } n = - \frac{1}{4}~,
 \nonumber \\  
 & \quad   \mbox{(only for transverse sectors)}~.
 \nonumber
\end{align}
Then, totally we get
\begin{align}
 & -a_B = -\frac{1}{3} - \half \omega ^2 + \half \omega ~.
 \label{eq:a_B_C}
\end{align}
It may be remarkable that the $\omega ^2$ anomaly appears in the bosonic sector.
%
\subsection{Fermionic sector}
%
The fermionic sector becomes
\begin{align}
 &  L_0^F = \half \sum_{r\geq 1/2} \big( r-\omega \big)  b_{-r}^{(+)}b_{r}^{(-)}
 \label{eq:L_0_F_8} \\ 
 & \quad +\half \sum_{r\geq 1/2} \big(- r-\omega \big)  b_{r}^{(+)}b_{-r}^{(-)} 
  \nonumber \\
 & \quad  +\half \sum_{r\geq 1/2}\big( r+\omega \big) b_{-r}^{(-)}b_{r}^{(+)}
 \nonumber \\
 & \quad + \half \sum_{r\geq 1/2}\Big(-r+\omega \big) b_{r}^{(-)}b_{-r}^{(+)} 
    \nonumber \\
 & \quad  +\half \sum_{r\geq 1/2}\sum_{\mu ,\nu \neq 1,2}\eta _{\mu \nu } r \big( b_{-r}^{\mu } b_{r}^{\nu } - b_{r}^{\mu } b_{-r}^{\nu } \big)
   \nonumber \\
 & = \sum_{r\geq 1/2} \big( r-\omega \big)  b_{-r}^{(+)}b_{r}^{(-)} + \half \sum_{r\geq 1/2} \big(- r-\omega \big)
   \nonumber \\
 &  \quad + \sum_{r\geq 1/2}\big( r+\omega \big) b_{-r}^{(-)}b_{r}^{(+)} + \half \sum_{r\geq 1/2} \big( -r+\omega \big)
   \nonumber \\
 & \quad + \sum_{r\geq 1/2}\sum_{\mu ,\nu \neq 1,2}\eta _{\mu \nu } r b_{-r}^{\mu } b_{r}^{\nu }
    - \half \sum_{r\geq 1/2}\sum_{\mu ,\nu \neq 1,2}\eta _{\mu \nu } \eta ^{\mu \nu } r~. 
   \nonumber
\end{align}
By using the formulae
\begin{align}
 & \half \sum_{r\geq 1/2} \big( -r-\omega \big) = -\frac{1}{48} + \frac{\omega ^2}{4}~,
\end{align}
\begin{align}
 & \half \sum_{r\geq 1/2} \big( -r+\omega \big) = -\frac{1}{48} + \frac{\omega ^2}{4}~,
 \label{eq:sum_C}
\end{align}
and
\begin{align}
 & -\half \sum_{r\geq 1/2}\sum_{\mu ,\nu \neq 1,2}\eta _{\mu \nu } \eta ^{\mu \nu } r =  -\frac{1}{8} 
 \nonumber \\
 & \quad \mbox{ (only for transverse sectors)}~,
\end{align}
we get
\begin{align}
 & - a_F = -\frac{1}{6} + \frac{\omega ^2}{2}~.
 \label{eq:a_F_C}
\end{align}
The total sum is given by
\begin{align}
 & a = a_B + a_F = \frac{1-\omega }{2}~.
 \label{eq:a_F_C2}
\end{align}
It is remarkable that the $\omega ^2$ anomaly is cancelled out in $a$. \\
\indent For the heterotic sring model we have
\begin{align}
 & a = \frac{1-\omega }{2}~,  
 \nonumber \\
 & \quad \mbox{for the right-moving NS sector}~,
 \label{eq:a_right-moving_C} \\
 & \tilde {a} = 1 - \frac{\omega -\omega^2 }{2}~, 
 \nonumber \\
 & \quad \mbox{for the left-moving sector}~.
 \label{eq:a_left-moving_C}   
\end{align}
%
%

\end{document}